\documentclass[namedreferences]{solarphysics}
\usepackage[optionalrh,solaromanenum]{spr-sola-addons} 
\usepackage{graphicx}                    
\usepackage{color}                       
\usepackage{url}                         

\usepackage[pdfborder={0 0 0},colorlinks=true,urlcolor=blue,breaklinks]
{hyperref}
\newcommand{\doiurl}[1]{\href{http://dx.doi.org/#1}{#1}}

\def\be{\begin{equation}}
\def\ee{\end{equation}}
\def\bea{\begin{eqnarray}}
\def\eea{\end{eqnarray}}

\newfont{\myfont}{cmmib10}

\def\bi{\bf}

\newcommand{\aap}{    {\it Astron. Astrophys.}}

\newcommand{\apj}{    {\it Astrophys. J.}}

\newcommand{\grl}{    {\it Geophys. Res. Lett.}}

\newcommand{\jgr}{    {\it J. Geophys. Res.}}
\newcommand{\mnras}{  {\it Mon. Not. Roy. Astron. Soc.}}

\newcommand{\solphys}{{\it Solar Phys.}}

\newcommand{\ssr}{    {\it Space Sci. Rev.}}

\begin{document}
\begin{article}
\begin{opening}
\title{%
Bulk Energization of Electrons in Solar Flares by Alfv\'en Waves
}
\author{%
D.B.~Melrose{}$^{1}$\sep M.S.~Wheatland{}$^{1}$
}
\runningauthor{D.B. Melrose, M.S. Wheatland}
\runningtitle{Bulk Energization by Alfv\'en Waves}

\institute{%
$^{1}$ Sydney Institute for Astronomy, School of Physics, University of Sydney, NSW 2006, Australia
email: \url{melrose@physics.usyd.edu.au,
m.wheatland@physics.usyd.edu.au} 
}

\begin{abstract}
Bulk energization of electrons to $10\,-\,20\,$keV in solar flares is attributed to dissipation of Alfv\'en waves that transport energy and potential downward to an acceleration region near the chromosphere. The acceleration involves the parallel electric field that develops in the limit of inertial Alfv\'en waves (IAWs). A two-potential model for IAWs is used to relate the parallel potential to the cross-field potential transported by the waves. We identify a maximum parallel potential in terms of a maximum current density that corresponds to the threshold for the onset of anomalous resistivity. This maximum is of order $10\,$kV when the threshold is that for the Buneman instability. We argue that this restricts the cross-field potential in an Alfv\'en wave to about $10\,$kV. Effective dissipation requires a large number of up- and down-current paths associated with multiple Alfv\'en waves. The electron acceleration occurs in localized, transient, anomalously-conducting regions (LTACRs) and is associated with the parallel electric field determined by Ohm's law with an anomalous resistivity. We introduce an idealized model in which the LTACRs are (upward-)current sheets, a few skin depths in thickness, separated by much-larger regions of weaker return current. We show that this model can account semi-quantitatively for bulk energization.
\end{abstract}

\end{opening}

\section{Introduction}
\label{s:introduction}
A long-standing problem in the physics of solar flares concerns the ``bulk energization'' or ``first phase acceleration'' of electrons. These electrons produce the characteristic observational features of the impulsive phase of a flare: hard X-ray bursts from footpoints of the flaring flux loops, and type~III radio bursts \cite{WSW}. Despite an extensive literature over many decades, no model has emerged that can account in a self-consistent way for important features inferred from observations. We summarize the difficulties in terms of three problems.

The first problem is the enormous energy channeled through these electrons: essentially all the energy in a non-eruptive flare and all the nonthermal energy in an eruptive flare go into electrons with kinetic energies $\gtrsim10\,$keV. The power transferred to these electrons can be written as $I\Phi$, with the current $I$ of order $10^{11}\,$A identified from vector magnetograph data, and one then requires a potential $\Phi$ of order $10^{10}\,$V. A simple estimate of the rate of change of the magnetic flux stored in the corona leads to a value of this order, but there is no direct evidence for such a large potential in the corona. The energy $\varepsilon$ of the accelerated electrons is of order $10^{4}\,$eV, and if this is identified as $e$ times a potential, the potential is of order $10^4\,$V. The problem is the inconsistency between the potential $\Phi\approx10^{10}\,$V required by the electrodynamics, and the potential seemingly available for acceleration, which is $\varepsilon/e=\Phi/M$, with $M$ of order $10^6$.

The second problem is the ``number problem'' \cite{Hetal76,MB89,BKB10} which can be expressed in several ways. One manifestation is that the number of electrons accelerated greatly exceeds the number of electrons in the flaring flux tube prior to the flare. This requires that the electrons be resupplied continuously by a return current \cite{BB84,SS84,vdO90,LS91,EH95}. The number problem can also be expressed in terms of the rate electrons precipitate, implied by a thick-target-bremsstrahlung model for hard X-ray bursts. This rate $\,{\dot{\!N}}$ exceeds $I/e$ by the same multiplicity factor $M$ of order $10^6$.

The third problem is that the acceleration mechanism must account for energization of all the electrons in a given volume, not just once but many times as the electrons are continuously resupplied, accelerated, and escape. Acceleration by shocks can lead to a form of bulk energization, but there is no evidence that the nonthermal energy in flares goes primarily into shocks. Acceleration by a parallel electric field $E_\parallel$ seems the only viable bulk-energization mechanism. Discussion of such acceleration in the older literature concentrated on the runaway effect, where effectively all the electrons experience runaway acceleration for $E_\parallel>E_{\rm D}$, where $E_{\rm D}$ is the Dreicer field. In the more recent literature, influenced by analogous problems that occur in the acceleration of auroral electrons in the Earth's magnetosphere, the energy transport into the acceleration region is assumed to be by Alfv\'en waves, and $E_\parallel\ne0$ is assumed to arise from dispersive effects associated with small-scale structures in these waves. This is the starting point for the discussion in the present article.

In the Alfv\'en-wave model that we adopt (\opencite{M12a}, \citeyear{M12b}, \opencite{H12}), the magnetic energy is stored in current-carrying loops in the corona prior to the flare. Magnetic reconnection allows transfer of magnetic flux and current to new magnetic loops, leading to a net reduction in the stored magnetic energy. The magnetic energy flows into a reconnection region where it is converted into an outflowing Alfv\'enic flux that transports the energy downward to an acceleration region near the chromosphere \cite{FH08}. In our version of this model \cite{MW13}, the Alfv\'enic flux involves transport of energy and potential, with the latter described by the cross-field potential $\Phi^+$ in a downgoing wave. The wave is assumed to damp in the acceleration region, transferring energy to accelerated electrons. An earlier attempt to model this acceleration on a macroscopic scale, in terms of a downward Poynting flux in the wave converting to a kinetic-energy flux in the accelerated electrons, led to an inconsistency \cite{MW13}. Although our starting assumption was a slow variation across field lines, the model leads to structure across field lines on a very small scale, determined by the skin depth $\lambda_{\rm c}$. This result further emphasizes the need for very small perpendicular scales for effective damping of Alfv\'en waves, as is widely recognized, both in connection with auroral acceleration (\opencite{G84}, \opencite{B93}, \opencite{K94}, \opencite{S00}, \opencite{Cetal03}, \citeyear{Cetal07}, \opencite{SL07}, \opencite{Aetal09}) and in connection with solar flares \cite{LM93,MF09,BKB10,H12}. Dispersive effects lead to Alfv\'en waves having $E_\parallel\ne0$, and the suggestion is that this $E_\parallel$ allows the required energy transfer from waves to particles. However, there is no consensus on the details of this mechanism. An older suggestion \cite{LM93} is that a turbulent cascade transfers the energy to small scales where Fermi acceleration transfers this energy to the electrons \cite{LMS94}. More recent suggestions \cite{MF09,BKB10} are that the small scales result in inertial Alfv\'en waves (IAWs), or in kinetic Alfv\'en waves (KAWs), that are damped due to electrons being accelerated by $E_\parallel$ in the wave.

In this article we explore a different approach to energy transfer from Alfv\'en waves to electrons. Any effective dissipation in a collisionless plasma, such as the solar corona, must involve some form of anomalous resistivity \cite{S82,H85}. Anomalous resistivity involves a current-driven plasma instability. When the parallel current density $J_\parallel$ exceeds an appropriate threshold $J_{\rm th}$ relevant waves grow, and the anomalous dissipation is due to the energy going into these waves being transferred to electrons, which damp them. We assume that such anomalous dissipation is the mechanism by which the energy in Alfv\'en waves is transferred to the electrons. The electric field $E_\parallel\ne0$ can be described in terms of the associated parallel potential $\Psi$, and we argue that this has a maximum ($\Psi_{\rm max}$) that implies a maximum energy $e\Psi_{\rm max}$ to which electrons can be accelerated. However, we do not attribute the acceleration of electrons to $E_\parallel\ne0$ in the waves. We assume that within the acceleration region, the accelerating electric field is determined by an Ohm's law ($E_\parallel\approx J_{\rm th}/\sigma_{\rm an}$) that involves the anomalous resistivity ($1/\sigma_{\rm an}$, where $\sigma_{\rm an}$ is the anomalous conductivity). We show that this model provides a natural explanation for the acceleration of the bulk of the electrons to a typical energy of tens of keV.

In Section~\ref{s:Alfven} we present a model for IAWs involving both a cross-field and a parallel potential. In Section~\ref{s:max} we show that there is a maximum parallel potential in the waves, determined by the threshold for exciting anomalous resistivity through a current-driven instability, and we argue that the maximum parallel potential is of order that required to account for bulk energization when the threshold corresponds to the Buneman instability. In Section~\ref{s:acc} we assume that anomalous resistivity develops in localized, transient, anomalously-conducting regions (LTACRs) and we outline a model for bulk energization of electrons in LTACRs. In Section~\ref{s:number} we present an idealized model in which the LTCARs are long thin upward-current sheets, separated by much larger regions of weaker return current, and we show how this model can account semi-quantitatively for bulk energization. We discuss our results in Section~\ref{s:discussion} and summarize our conclusions in Section~\ref{s:conclusions}.

\section{Inertial Alfv\'en Waves}
\label{s:Alfven}
In this section we first comment on dispersive, inertial, and kinetic Alfv\'en waves. We then extend a treatment of large-amplitude Alfv\'en waves \cite{MW13} in terms of the wave potential $\Phi$ to include the parallel potential $\Psi$ in the waves.

\subsection{DAWs, IAWs, and KAWs}
Alfv\'en waves in the approximation $E_\parallel=0$ are nondispersive, and when $E_\parallel\ne0$ is included they are referred to as dispersive Alfv\'en waves (DAWs). Two limiting cases of DAWs are inertial Alfv\'en waves (IAWs) and kinetic Alfv\'en waves (KAWs)
\cite{H76,LL96,S00,LS03,DWM09}. The dispersion relation and ratio of parallel to perpendicular electric field for IAWs and KAWs are
\be
\omega^2={k_z^2v_0^2\over1+\lambda_{\rm c}^2k_\perp^2},
\qquad
{E_\parallel\over E_\perp}={k_\perp\over k_z}{\lambda_{\rm c}^2k_\perp^2\over1+\lambda_{\rm c}^2k_\perp^2},
\label{IAW}
\ee
and
\be
\omega^2=k_z^2v_{\rm A}^2(1+R_{\rm g}^2k_\perp^2),
\qquad
{E_\parallel\over E_\perp}=-{k_\perp\over k_z}R_{\rm g}^2k_\perp^2,
\label{KAW}
\ee
respectively. In Equation (\ref{IAW}), $\omega$, $k_z$, and $k_\perp$ are the frequency, parallel wavenumber, and perpendicular wavenumber, respectively, of the wave, $v_0=v_{\rm A}/(1+v_{\rm A}^2/c^2)^{1/2}$ is the MHD speed, with $v_{\rm A}$ the Alfv\'en speed, and $\lambda_{\rm c}=c/\omega_{\rm p}$ is the skin depth with $\omega_{\rm p}=(e^2n_{\rm e}/\varepsilon_0m_{\rm e})^{1/2}$ the plasma frequency. An explicit expression for $R_{\rm g}$ in Equation (\ref{KAW}) is given in Equation (\ref{Rg}) in Appendix~\ref{A-appendix}.

In a derivation from kinetic theory \cite{LS03}, Equation (\ref{IAW}) follows when the electrons are assumed cold ($\omega^2\gg k_z^2V_e^2$) and Equation (\ref{KAW}) follows when the opposite approximation ($\omega^2\ll k_z^2V_{\rm e}^2$) is made for the electrons. These limits correspond to $V_{\rm e}^2\ll v_0^2$ and $v_{\rm A}^2\ll V_{\rm e}^2\ll c^2$, respectively, and in this model $R_{\rm g}=\lambda_{\rm c}V_{\rm e}/v_{\rm A}\gg\lambda_{\rm c}$. If follows that IAWs and KAWs are opposite limiting cases and that only one of them is relevant in a given plasma. The following discussion applies to IAWs, which are the waves of relevance in the corona, where one has $V_{\rm e}\approx10^6\rm\,m\,s^{-1}$ and $v_{\rm A}\approx10^7\rm\,m\,s^{-1}$. KAWs are discussed in Appendix~\ref{A-appendix}.

The presence of $E_\parallel\ne0$ allows Landau damping of the waves when the resonance condition
\be
\omega-k_zv_z=0
\label{Landau}
\ee
is satisfied, where $v_z$ is the parallel component of the resonant particle. For IAWs this requires $v_z>v_0\gg V_{\rm e}$, and for KAWs it requires $v_z<v_{\rm A}\ll V_{\rm e}$. In neither case does Landau damping allow acceleration of thermal electrons, with $v_z\approx V_{\rm e}$, as required for bulk energization.

\subsection{Two-potential Model for DAWs}
The electric field in an Alfv\'en wave may be described in terms of two potentials \cite{C05}, one ($\Phi$) corresponding to the perpendicular electric field, and the other ($\Psi$) corresponding to the parallel electric field:
\be
{\bi E}_\perp=-{\bf\nabla}_\perp\Phi,
\qquad
E_\parallel=-{\partial\Psi\over\partial z}.
\label{2pot}
\ee
The potentials satisfy two coupled equations, derived from the wave equation, which follows from Maxwell's equations including the response of the plasma through the induced current density ${\bi J}$. After separating into perpendicular and parallel components and assuming that the perpendicular response is due to the polarization current \cite{MW13}, one finds
\be
{1\over v_0^2}{\partial^2{\bi E}_\perp\over\partial t^2}-\nabla^2{\bi E}_\perp
+{\bf\nabla}_\perp({\bf\nabla}_\perp\cdot{\bi E}_\perp)
=-{\bf\nabla}_\perp{\partial E_\parallel\over\partial z},
\label{rc1a}
\ee
\be
{1\over c^2}{\partial^2E_\parallel\over\partial t^2}-\nabla_\perp^2E_\parallel
+{\partial\over\partial z}({\bf\nabla}_\perp\cdot{\bi E}_\perp)
=-\mu_0{\partial J_\parallel\over\partial t}.
\label{rc1b}
\ee
The component of Equation (\ref{rc1a}) that corresponds to ${\bf\nabla}_\perp\times{\bi E}_\perp\ne0$ describes magnetoacoustic waves, and is ignored here. Inserting Equation (\ref{2pot}) into Equation (\ref{rc1a}) gives one of the two coupled equations for Alfv\'en waves:
\be
\left[{1\over v_0^2}{\partial^2\over\partial t^2}-{\partial^2\over\partial z^2}\right]\Phi
=-{\partial^2\Psi\over\partial z^2}.
\label{we1}
\ee
Applying the same procedure to Equation (\ref{rc1b}) gives
\be
{\partial\over\partial z}\left\{\left[{1\over c^2}{\partial^2\over\partial t^2}-\nabla_\perp^2\right]\Psi+\nabla_\perp^2\Phi\right\}
=\mu_0{\partial J_\parallel\over\partial t}.
\label{we2}
\ee
A model for the induced parallel current density $J_\parallel$ is required, and this is different for IAWs and KAWs.

A more conventional description of the wave fields, in terms of scalar and vector potentials, was used by \inlinecite{S00} to derive the properties of IAWs. In Appendix~\ref{B-appendix} we show that the two descriptions are related by a gauge transformation.

\subsection{Wave Equation for IAWs}
IAWs apply in the limit where the electrons are treated as cold. A nonrelativistic electron obeys Newton's equation in the form
\be
m_{\rm e}{dv_z\over dt}=-eE_\parallel-\nu_{\rm eff}m_{\rm e}v_z,
\label{IAW1}
\ee
where $m_{\rm e}$ and $-e$ are the mass and charge of the electron, respectively, and an effective collision frequency $\nu_{\rm eff}$ is included for later purposes. Assuming $J_\parallel=-en_{\rm e}v_z$, where $n_{\rm e}$ is the electron number density, Equation (\ref{IAW1}) gives
\be
{dJ_\parallel\over dt}=\varepsilon_0\omega_{\rm p}^2E_\parallel-\nu_{\rm eff}J_\parallel.
\label{IAW2}
\ee
Inserting Equation (\ref{IAW2}) into Equation (\ref{we2}) gives
\be
{\partial\over\partial z}\left\{\left[{1\over c^2}{\partial^2\over\partial t^2}
+{\omega_{\rm p}^2\over c^2}
-\nabla_\perp^2\right]\Psi+\nabla_\perp^2\Phi\right\}
=-\mu_0\nu_{\rm eff} J_\parallel.
\label{IAW3}
\ee
The second of the coupled wave equations for IAWs follows from Equation (\ref{IAW3}) by assuming that the wave frequency is much smaller than $\omega_{\rm p}$, justifying neglect of the term involving $\partial^2/\partial t^2$. We also neglect the right hand side (except within LTACRs). This gives
\be
(1-\lambda_{\rm c}^2\nabla_\perp^2)\Psi+\lambda_{\rm c}^2\nabla_\perp^2\Phi=0.
\label{IAW4}
\ee
Combining Equations (\ref{we1}) and (\ref{IAW4}), the wave equation for IAWs becomes
\be
\left[{(1-\lambda_{\rm c}^2\nabla_\perp^2)\over v_0^2}{\partial^2\over\partial t^2}-{\partial^2\over\partial z^2}\right]
\left(
\begin{array}{c}\Phi\\
\Psi
\end{array}
\right)
=0.
\label{IAW5}
\ee
After Fourier transforming, Equations (\ref{2pot}), (\ref{IAW4}), and (\ref{IAW5}) imply the wave properties in Equation (\ref{IAW}).

\section{Restriction on the Parallel Potential}
\label{s:max}
In this section we argue that the two-potential model leads to an estimate of the maximum energy $\varepsilon_{\rm max}=e\Psi_{\rm max}$ to which electrons can be accelerated by $E_\parallel$. We identify $\varepsilon_{\rm max}$ as the characteristic energy associated with bulk energization.

\subsection{Solution of the Wave Equation}
In the absence of dispersion, the wave equation [Equation (\ref{IAW5})] with $\lambda_{\rm c}\to0$ has solutions of the form $\Phi^\pm(x,y,z\pm v_0t)$, corresponding to downward- and upward-propagating waves, respectively. On Fourier transforming in $x$ and $y$, the operator $1-\lambda_{\rm c}^2\nabla_\perp^2$ in Equation (\ref{IAW5}) is replaced by $1+\lambda_{\rm c}^2k_\perp^2$, with $k_\perp^2=k_x^2+k_y^2$. The downward- and upward-propagating waves can then be described by ${\tilde\Phi}^\pm(k_x,k_y,z\pm v_\phi t)$ and ${\tilde\Psi}^\pm(k_x,k_y,z\pm v_\phi t)$, where the tilde denotes Fourier transforming in $x$ and $y$, and where
\be
v_\phi=v_0(1+\lambda_{\rm c}^2k_\perp^2)^{1/2}
\label{vphi}
\ee
is the phase velocity of the waves along the field lines. Equation (16) of \inlinecite{MW13}, which gives the parallel current density in the wave, is then replaced by
\be
{\tilde J}'^\pm_\parallel={\tilde J}^\pm_\parallel-\varepsilon_0{\partial^2{\tilde\Psi}^\pm\over\partial t\partial z}=
\mp {k_\perp^2{\tilde\Phi}^\pm\over\mu_0v_\phi},
\label{Jpar1}
\ee
where arguments are omitted for simplicity in writing. Using Equation (\ref{IAW4}), one finds
\be
{\tilde J}^\pm_\parallel=\mp{k_\perp^2\over R_{\rm A}(1+k_\perp^2\lambda_{\rm c}^2)^{1/2}}
\left[1-\lambda_{\rm c}^2{v_0^2\over c^2}{\partial^2\over\partial z^2}
\right]{\tilde\Phi}^\pm,
\label{Jpar2}
\ee
where $R_{\rm A}=\mu_0v_0$ is the Alfv\'enic impedance. The second term inside the square brackets in Equation (\ref{Jpar2}) is due to the parallel displacement current, and can be neglected for $v_0^2\ll c^2$. Using Equation (\ref{IAW4}) again, one finds a relation between the parallel potential and the parallel current density:
\be
{\tilde\Psi}^\pm=\mp {\lambda_c^2\over(1+k_\perp^2\lambda_{\rm c}^2)^{3/2}}\,R_{\rm A}{\tilde J}^\pm_\parallel.
\label{Jpar3}
\ee
On expanding in powers of $k_\perp^2\lambda_{\rm c}^2$ and inverting the Fourier transform, Equation (\ref{Jpar3}) gives
\be
\Psi^\pm=\mp \lambda_{\rm c}^2R_{\rm A}\left(1+{3\over2}\lambda_{\rm c}^2\nabla_\perp^2+\cdots\right)J^\pm_\parallel.
\label{Jpar4}
\ee
For semi-quantitative purposes we approximate Equation (\ref{Jpar4}) by retaining only the unit term inside the round brackets, and this is valid provided that the characteristic length over which wave fields vary across field lines is not significantly smaller than $\lambda_{\rm c}$.

We emphasize that by Alfv\'en `wave' we mean a solution of the relevant Alfv\'en wave equation, and that the Alfv\'en waves of relevance here are not periodic in space and time. Our solutions correspond to the long-wavelength limit, with the (ill-defined) wavelength of order the distance between the generator and acceleration regions. All quantities associated with the wave are slowly varying functions of $z$ and $t$. If we were to apply our model to a plane wave, all the wave quantities would oscillate periodically, including the (parallel) energy of particles, which would oscillate with an amplitude $e\Psi^\pm$. For net particle acceleration to result, some form of dissipation is required to introduce time-irreversibility into the model. This remains the case in the long-wavelength limit.

The parallel potential in the IAW is proportional to the parallel current density, which varies across the field lines. Effective energy transport occurs when the direct current is strongly concentrated in a small central region, surrounded by a larger region with a weaker return current \cite{M12b,MW13}. It follows from Equation (\ref{Jpar4}) that the parallel potential has a similar profile, being large in the small central region of direct current, and smaller and of the opposite sign in the surrounding region of return current. The maximum parallel potential $\Psi_{\rm max}$ is determined by the maximum current density, which is in the small central region of direct current.

\subsection{Maximum Parallel Potential}
There is a maximum parallel current density ($J_\parallel=J_{\rm max}$) that can flow in a plasma before a current-driven instability turns on. Let the threshold for instability correspond to
\be
J_{\rm th}=en_{\rm e}v_{\rm th},
\label{Jth}
\ee
where $v_{\rm th}$ is the threshold velocity for the instability. The backreaction to the development of the instability tends to reduce $J_\parallel$ to $J_{\rm th}$. We equate $J_{\rm max}$ to $J_{\rm th}$ for semi-quantitative purposes. It follows from Equation (\ref{Jpar4}) that there is a maximum parallel potential, and a maximum energy to which electrons can be accelerated by $E_\parallel$:
\be
\varepsilon_{\rm max}=e\Psi_{\rm max}=
e\lambda_{\rm c}^2R_AJ_{\rm max}=m_{\rm e}v_{\rm A}v_{\rm th}
=T_{\rm e}{v_{\rm A}v_{\rm th}\over V_{\rm e}^2},
\label{emax}
\ee
where $T_{\rm e}=m_{\rm e}V_{\rm e}^2$ is the electron temperature (in energy units).

We regard Equation (\ref{emax}) as one of the important results of this article. Qualitatively, it shows that the model cannot result in acceleration to the energy $e\Phi$, as assumed by \inlinecite{C78} for example, and it restricts the maximum energy to a much smaller value. Semi-quantitatively, the value given by Equation (\ref{emax}) with $v_{\rm A}\approx10\,V_{\rm e}$ and $v_{\rm th}$ a few times $V_{\rm e}$ can plausibly account for the energy $10\,-\,20\,$keV inferred from observations.

Equations (\ref{Jth}) and (\ref{emax}) may be used to rewrite the threshold condition ($J_\parallel=J_{\rm th}$) for instability. The parallel current density and parallel potential in downward Alfv\'en waves may be written in the form
\be
{J_\parallel^+\over J_{\rm th}}={\Psi^+\over\Psi_{\rm max}}
\approx-{\lambda_{\rm c}^2\nabla_\perp^2\Phi^+\over\Psi_{\rm max}},
\label{thresh}
\ee
where Equation (\ref{IAW4}) is used in the approximate relation. Instability develops along the current path when the ratio given by Equation (\ref{thresh}) exceeds unity. Introducing a perpendicular scale length $\lambda_\perp=2\pi/k_\perp$ by $\nabla_\perp^2\to-k_\perp^2$, this favors regions where $k_\perp\lambda_{\rm c}$ is largest.

\subsection{Buneman Instability}
We apply Equation (\ref{emax}) to the Buneman instability (\opencite{B58}, \citeyear{B59}), which has a threshold $v_{\rm th}\approx V_{\rm e}$. The Buneman instability is reactive, in the sense that the wave growth involves a complex solution of a real dispersion equation \cite{M86}. The reactive nature applies only when the electrons can be regarded as cold, and this applies only if the electrons have a drift velocity that exceeds their thermal spread, of order $V_{\rm e}$. The backreaction to the development of Buneman instability tends both to reduce the drift speed and to increase the thermal spread. This increase in thermal spread corresponds to a bulk energization, that is, it corresponds to an increase in $T_{\rm e}$. This heating effect is well known in the context of anomalous resistivity \cite{BE05}, where the Buneman instability is invoked to provide the electron heating (to $T_{\rm e}\gg T_{\rm i}$ where $T_{\rm i}$ is the ion temperature) required for the ion-acoustic instability to develop, or for double layers to form.

The maximum energy implied by Equation (\ref{emax}) with $v_{\rm A}/V_{\rm e}\gg1$ in an IAW, and with $v_{\rm th}/V_{\rm e}>1$ for the Buneman instability, can plausibly account for the electron energies inferred in bulk energization. Equation (\ref{emax}) then implies that $\varepsilon_{\rm max}$ exceeds the thermal energy $T_{\rm e}$ by a factor of order $v_{\rm A}/V_{\rm e}$, which is of order 10. The existence of this maximum is more important than its actual value, which is subject to uncertainties associated with the values of both $T_{\rm e}$ and $v_{\rm A}$. Both $T_{\rm e}$ and $v_{\rm A}$ are affected by the acceleration, with $T_{\rm e}$ increasing as a result of bulk energization, and $v_{\rm A}$ increasing due to the formation of a density cavity, as discussed below.

\section{Electron Acceleration in LTACRs}
\label{s:acc}
The bulk energization mechanism we propose here involves energy being transported into the acceleration region by a large-amplitude Alfv\'en wave and dissipated in a current sheet. The concentration of current (into sheets) is required for sufficient energy to be transported \cite{M12b,MW13}. The Buneman instability is assumed to develop within the sheets and produce anomalous resistivity. We refer to such regions as LTACRs. We discuss three aspects of the acceleration within LTACRs: the role of anomalous resistivity, energy flow into individual LTACRs, and the number of LTACRs required to be accelerating electrons at any given time.

\subsection{Role of Anomalous Resistivity}
The parallel electric field in an IAW does not lead automatically to energy dissipation and associated electron acceleration. In a homogeneous medium, an IAW with given $k_z$ and $k_\perp$ has a non-zero $E_\parallel$ that oscillates in time and space, and is associated with periodic recycling of energy within the wave.

Electron-ion collisions lead to dissipation, due to the energy associated with the forced motion of an electron in the wave being lost when it is randomized during a collision. This effect may be modeled by the drag term in the equation of motion [Equation (\ref{IAW1})], with $\nu_{\rm eff}$ then identified as the collision frequency. This dissipation may be described in terms of a resistivity $\nu_{\rm eff}/\varepsilon_0\omega_{\rm p}^2$ or a conductivity
\be
\sigma_{\rm an}={\varepsilon_0\omega_{\rm p}^2\over\nu_{\rm eff}}.
\label{an1}
\ee
The simplest model for anomalous resistivity or conductivity involves interpreting $\nu_{\rm eff}$ as an effective collision frequency that is very much greater than the electron-ion collision frequency.

The inclusion of dissipation, through the anomalous conductivity, implies that the electric field has both reactive (time-reversible) and dissipative (time-irreversible) parts. For example, when the response of the plasma is described in terms of the equivalent dielectric tensor, these parts of the response are due to the Hermitian and anti-Hermitian parts of the tensor, respectively. In the case of weak dissipation these two parts are independent of, and proportional to $\sigma_{\rm an}$, respectively. Any dissipation causes the amplitude of the wave to decrease, with the energy lost by the wave transferred (irreversibly) to the particles. In the long-wavelength limit the dissipative response of the plasma leads to a quasi-static $E_\parallel\ne0$. Here, we attribute the acceleration of the particles to the quasi-static $E_\parallel\ne0$ in the anomalously conducting region.

We assume that the parallel electric field that is important in the dissipation of the Alfv\'en waves results from the current density and the anomalous resistivity:
\be
E_\parallel={J_\parallel\over\sigma_{\rm an}}={\nu_{\rm eff}J_\parallel\over\varepsilon_0\omega_{\rm p}^2}.
\label{an2}
\ee
Further assuming that the current density is at the threshold of instability, Equation (\ref{an2}) determines $E_\parallel$ within a LTACR. Let the LTACR be of length $\ell_\parallel$ along the field lines, where $\ell_\parallel$ may be the length of a single region or the sum of the length of several regions. The total potential drop $E_\parallel\ell_\parallel$ cannot exceed the maximum parallel potential in the wave. Assuming an equality, this condition gives
\be
\nu_{\rm eff}\ell_\parallel=v_{\rm A},
\label{an3}
\ee
where we assume $v_{\rm A}\ll c$. Equation (\ref{an3}) was derived by \inlinecite{H12}. With this assumption, a typical electron in the dissipation region is accelerated to $\varepsilon_{\max}$, given by Equation (\ref{emax}), before escaping from the region.

\subsection{Energy Inflow into the Dissipation Region}
\label{sect:energy_flow}
Dissipation is confined to individual LTACRs, and energy must flow into a LTACR in order to be dissipated. We discuss three contributions to this energy inflow: energy flow along field lies, energy flow across field lines, and motion of LTACRs across field lines.

The energy flux in an Alfv\'en wave is given by the Poynting vector ${\bi S}^+\!={\bi E}^+\!\times{\bi B}^+\!/\mu_0$. In the non-dispersive limit the energy flux is parallel to the (background) magnetic field; $S_z^+\!=v_0(B_\perp^+)^2/\mu_0$, ${\bi S}_\perp^+=0$. One contribution to the energy flow into a LTACR is from $S_z^+$ along the field lines that pass through the LTACR. Although $S_z^+$ has maxima near the edges of the LTACR, much of the Poynting flux $S_z^+$ flows along field lines that do not enter the LTACR through its top. Effective dissipation requires that all the energy flows into the dissipation region. We identify two ways in which energy can flow sideways into a LTACR.

When $E_\parallel^+\ne0$ is taken into account in the IAW, the Poynting vector has a component across the field lines, with $S_\perp^+/S_z^+=-E_\parallel^+/E_\perp^+$. With $S_\perp^+\ne0$ the energy flux in the Alfv\'en waves can flow into a LTACR through one side.

In the discussion so far we neglect the fact that the presence of $E_\parallel\ne0$ implies that the frozen-in condition is not valid. At the top of the acceleration region, plasma moves with velocity ${\bi u}^+\!={\bi E}_\perp^+\times{\hat{\bi z}}/B_0$, where $B_0$ is the background magnetic field. The velocity ${\bi u}^+\!$ reduces to zero at the bottom of the acceleration region. This may be attributed to the slippage of plasma relative to field lines, and it can be modeled by assuming that the LTACRs are aligned at a slightly oblique angle to the vertical \cite{H12}. This model allows a vertically downward energy flux to enter a LTACR through its side. In Section~\ref{sect:stat} we describe a ``picket-fence'' model \cite{MW13} with a slant (along the $x$-axis) of the pickets containing the LTACRs. A strictly downward energy flux necessarily encounters a dissipation region if the slope is such that the top of one (slanted) dissipation region lies vertically above the bottom of the neighboring dissipation region. Figure~\ref{fig:MW2} shows the proposed geometry (the details of the model are discussed in Section~\ref{sect:stat}).

Assuming that the ray path of the Alfv\'en waves always encounters at least one LTACR, the rate of energy inflow into the LTACR can be estimated semi-quantitatively. We assume that the power inflow is
\be
P_{\rm in}= A_{\rm an}\lambda_{\rm c}^2{\ell_\parallel\nu_{\rm eff}\over v_{\rm A}}R_{\rm A}J_{\rm max}^2,
\label{an4}
\ee
where $A_{\rm an}$ is the surface area over which the current flows into the LTACR. In a planar model of a LTACR, the current is confined to a sheet of thickness $\ell_x$ and length $\ell_y\gg\ell_x$, and the magnetic field due to the current is $B_y\approx\mu_0J_{\rm max}\ell_x$ near the edge of the sheet. The power flowing along the $x$-axis into the LTACR is $P_{\rm in}=\ell_y\ell_\parallel E_\parallel B_y/\mu_0$, which reproduces Equation (\ref{an4}) with $A_{\rm an}=\ell_x\ell_y$. The total current flowing in the LTACR is $I_{\rm max}=J_{\rm max}A_{\rm an}$, and Equation (\ref{an4}) may be written as
\be
P_{\rm in}= {\lambda_{\rm c}^2\over A_{\rm an}}R_{\rm A}I_{\rm max}^2,
\label{an5}
\ee
where Equation (\ref{an3}) is assumed to apply.

The energy flowing into a LTACR, given by Equation (\ref{an5}), must be balanced by the power dissipated within the LTACR. The power dissipated per unit volume is $E_\parallel J_\parallel=J_{\rm max}^2/\sigma_{\rm an}$. On multiplying by the volume $A_{\rm an}\ell_\parallel$ of the LTACR, in which the dissipation is confined, this gives
\be
P_{\rm diss}=A_{\rm an}\ell_\parallel\,{J_{\rm max}^2\over\sigma_{\rm an}}
=R_{\rm eff}I_{\rm max}^2,
\qquad
R_{\rm eff}={\lambda_{\rm c}^2\over A_{\rm an}}R_{\rm A},
\label{an6}
\ee
where Equation (\ref{an3}) is assumed to apply. The equality $P_{\rm in}=P_{\rm diss}$ is a self-consistency constraint on the model. Equation (\ref{an6}) may be rewritten in the form
\be
P_{\rm diss}=\Psi_{\rm max}I_{\rm max},
\qquad
\Psi_{\rm max}=R_{\rm eff}I_{\rm max},
\label{an7}
\ee
with $R_{\rm eff}$ interpreted as the resistance of the LTACR.

\subsection{Conservation of Current and Charge within LTACRs}
At the top of the acceleration region, $\Phi^+$, $\Psi^+$, and $J_\parallel^+$ are all associated with the downward-propagating Alfv\'en wave which transports energy. Assuming that all the energy is transferred to electrons within the acceleration region, one has $\Phi^+=0$ and $\Psi^+=0$ below a LTACR, but $J_\parallel$ is not zero in this region. This leads to several related problems that need to be addressed in a more detailed model for the acceleration region.

\begin{itemize}
\item {\it Conservation of current}:
The parallel current at the top of the acceleration region is $J_\parallel^+$ associated with the Aflv\'en wave, and both $J_\parallel^+$ and $J_\perp^+$ must decrease in magnitude as the wave damps. The electric field $E_\parallel$ in a LTACR accelerates electrons, tending to increase $J_\parallel$. Conservation of current requires that this new contribution to $J_\parallel$ be balanced by an associated $J_\perp$.
\item {\it Conservation of charge}: There is a net outflow of accelerated electrons from the LTACR. To conserve charge, either electrons are resupplied continuously to the LTACR, or ions are lost at the same rate as electrons are lost. Outflow of ions across field lines as a polarization current associated with a (temporally) changing ${\bi E}_\perp$ seems the only plausible mechanism.
\item {\it Current closure}:
The current $J_\parallel$ and associated $J_\perp$ that arise in association with the acceleration of electrons must be part of a new closed current loop. Where this new current closes needs to be identified. We assume that the current loop involves field-aligned currents to and from a conducting boundary, as proposed by \inlinecite{S55} in a laboratory context. A plausible region where cross-field current closure is possible is the partially-ionized region (near the photosphere), where a Pedersen current can flow.
\item {\it Cavity formation}:
Assuming that the increasing $J_\parallel$ associated with acceleration of electrons is balanced by a $J_\perp$ due to a polarization current carried by ions, the new closed current loop causes depletion of plasma within the LTACR. This cannot continue indefinitely, and either the LTACR dies away, and (statistically) is replaced by another LTACR developing in a higher density region, or the LTACR propagates across field lines leaving the underdense plasma in its wake.
\end{itemize}

\section{The ``Number Problem''}
\label{s:number}
In this section we discuss implications of our model on the long-standing number problem. We argue that our identification of the maximum parallel potential $\Psi_{\rm max}$ is an important new step in resolving the number problem, and we outline a model that demonstrates how resolution can be achieved.

\subsection{Link between Global- and Micro-scale Processes}
The number problem can be quantified by the requirement that the (nonthermal) power $I\Phi$ in a flare appears in electrons accelerated at a rate $\,{\dot{\!N}}=MI/e$ with energy $\varepsilon=e\Phi/M$, with $M=10^6$ for the fiducial numbers $I=10^{11}\,$A and $\Phi=10^{10}\,$V adopted here. Any acceptable resolution of the number problem requires an explanation for the multiplicity $M$. Following a suggestion by \inlinecite{H85}, we proposed the picket-fence model \cite{MW13}, in which the current $I$ flows up and down $M$ times between the generator and acceleration regions.

An important additional feature that we include here is the existence of $\Psi_{\rm max}$, which is given by Equation (\ref{emax}), and which is plausibly of order $10^4\,$V, allowing it to be equated to $\Phi/M$. This provides a way of explaining $M$, as the ratio $\Phi/\Psi_{\rm max}$ of the EMF generated by the changing magnetic field on a global scale, to the maximum parallel potential that the microphysics allows locally in the acceleration region. Another aspect of the link between these two scales concerns $J_\parallel$. The global electrodymanics requires that the Alfv\'en waves transport a large power, and the power is maximized by having the direct $J_\parallel$ concentrated in a very narrow channel surrounded by a much larger region of weaker return current \cite{M12b}. The microphysics requires that, for anomalous resistivity to develop, $J_\parallel$ must exceed the threshold for a current-driven instability. This effect is already included in the model, through $\Psi^+\propto J_\parallel^+$ in the IAW allowing $J_{\rm max}$ to determine $\Psi_{\rm max}$.

\subsection{Multiplicity}
We assume that prior to a flare the stressed magnetic field is unable to relax, and that the trigger for a flare is the turning on of anomalous dissipation. This causes the frozen-in condition to break down, allowing the magnetic field to slip relative to the plasma and allowing the magnetic field on a global scale to start to change. The changing magnetic field implies $\Phi\ne0$, which acts as the EMF that drives the current $I$ across field lines in the generator region. The cross-field $I$ and $\Phi$ launch an Alfv\'en wave that sets up a current loop that redirects $I$ along field lines through the dissipation region, where the energy is dissipated. The restriction of the parallel potential to $\Psi_{\rm max}$ in the dissipation region implies that the rate of change of magnetic flux on the global scale is restricted to a value $\Delta\Phi=\Psi_{\rm max}$. The now unbalanced magnetic stress drives the changes faster, and this can be achieved only by the current being redirected through the dissipation region multiple times. This results in the proposed picket-fence model, in which the current is redirected up and down $M$ times, allowing the rate of change of magnetic flux to increase to $\Phi=M\Delta\Phi=M\Psi_{\rm max}$.

In this model, there is a resistance $R_{\rm eff}$, given by Equation (\ref{an6}), associated with each section of the current path through the dissipation region. These resistances are in series along the current path, so that the total resistance is $MR_{\rm eff}$. With our fiducial numbers, the total effective resistance is equal to $\Phi/I=0.1\rm\,ohm$. Using Equations (\ref{an6}) and (\ref{an3}), this leads to a relation between the global electrodynamics and the properties of the LTACRs on the micro-scale:
\be
M={A_{\rm an}\over\lambda_{\rm c}^2}{I\Phi\over R_{\rm A}I_{\rm max}^2}
={A_{\rm an}\over\lambda_{\rm c}^2}{I^2\over I_{\rm max}^2}{\Phi\over R_{\rm A}I}.
\label{npa}
\ee
Assuming $v_{\rm A}=10^7\rm\,m\,s^{-1}$, which implies $R_{\rm A}\approx10\,$ohm, and also assuming $I_{\rm max}=I$, our fiducial numbers require $A_{\rm an}/\lambda_{\rm c}^2$ of order $10^8$. Thus we find that the global requirements can be met by the microphysics, in a model based on many LTACRs, provided that the area associated with each LTACR is of order $10^8\lambda_{\rm c}^2$.

\begin{figure} [t]
\centerline{
\includegraphics[scale=0.5]{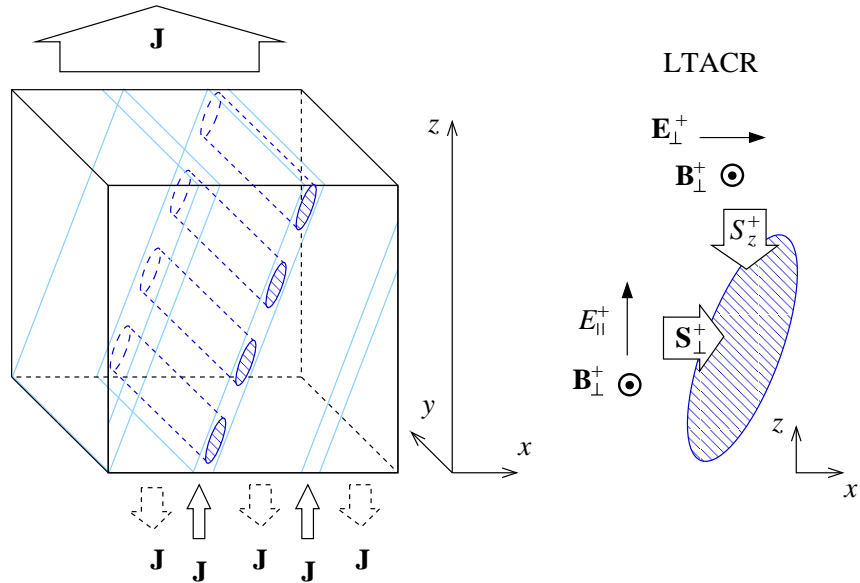}
}
\caption{%
The model for LTACRs is illustrated schematically. The large arrow at the top of the figure on the left indicates the direction of the average current, and the solid and dashed arrows at the bottom indicate the direct current, confined to thin sheets indicated by the faint lines, and the return current between these sheets. The regions of anomalous resistivity are indicated in one sheet by cylinders along the $y$-axis with elliptical cross-section with major axis along the $z$-axis. The figure on the right indicates the direction of the parallel electric field and the energy flow in the IAW at the surface of a LTACR. The sheets are shown slanted to take account of the slippage of plasma relative to magnetic field.
}
\label{fig:MW2}
\end{figure}

\subsection{Statistical Model for the Acceleration Region}
\label{sect:stat}
A simplifying feature of a model for anomalous dissipation based on a large number of LTACRs is that the macroscopic effects of the dissipation depend on the distribution of LTACRs, and are relatively insensitive to the detailed properties of individual LTACRs. We suggest a model in which the LTACRs are idealized as current sheets that extend across the flaring flux tube in one direction (the $y$-axis) and correspond to the picket-fence structure in the other direction (the $x$-axis). This model is illustrated in Figure~\ref{fig:MW2}, which shows three of the $M$ sheets along the $x$-axis, each of which includes four LTACRs. The maximum current density $J_\parallel=J_{\rm max}$ is confined to sheets of length $\ell_y$ and thickness $\ell_x$, separated by regions of return current of thickness $\gg\ell_x$. The model is not sensitive to the distribution of anomalous conductivity either along the field lines or along the $y$-axis. Along the field lines, any number of LTACRs with a net length $\ell_\parallel$ is equivalent to a single LTACR of length $\ell_\parallel$. The distribution of LTACRs along the $x$-axis is highly structured. There are $M$ current sheets and the LTACRs are confined to these sheets. The direct current flows within these sheets, which have only a small filling factor along the $x$-axis, with the much larger regions between the sheets carrying the much weaker return current.

The current sheets illustrated in Figure~\ref{fig:MW2} are shown with an (exaggerated) slant. A slant results from the slippage of the plasma relative to the magnetic field \cite{H12}, and is one of the mechanisms (discussed in Section~\ref{sect:energy_flow}) that allows all the downward energy flux to encounter a LTACR where the energy is dissipated. The panel on the right in Figure~\ref{fig:MW2} indicates how both the slant and the horizontal component of the Poynting flux can result in the downward energy flux entering a LTACR through its side.

An important parameter in this model is the area $A_{\rm an}=\ell_x\ell_y$ of a LTACR, which also determines the filling factor of the regions with $J_\parallel=J_{\rm max}$ along the $x$-axis. This area $A_{\rm an}/(\Delta R)^2$, where $\Delta R$ is the width of the flux tube, is related to the multiplicity $M$ through Equation (\ref{an6}). For our model to account quantitatively for bulk energization we require $A_{\rm an}\approx10^8\lambda_{\rm c}^2$. To see how this can be achieved, first note that to achieve the maximum parallel potential in an IAW requires $k_\perp\lambda_{\rm c}\approx1$, which corresponds to $\ell_x\approx2\pi\lambda_{\rm c}$. With $\ell_y=\Delta R$ of order $10^6\,$m, one has $\Delta R=10^7\lambda_{\rm c}$ for $\lambda_{\rm c}$ of order $10^{-1}\,$m corresponding to an electron density $n_{\rm e}=10^{16}\rm\,m^{-3}$. With these numbers, the filling factor of the regions with $J_\parallel\approx J_{\rm max}$ is of order $10^{-2}$.

A self-consistency check on the model is given by estimating the total rate electrons are accelerated. We assume that electrons are accelerated to a speed $v_{\rm esc}=(2\varepsilon_{\rm max}m_{\rm e})^{1/2}$ before they escape from an individual LTACR. The rate electrons escape from a LTACR is then
\be
{\,\dot{\!N}}_{\rm c}=(n_{\rm e}\lambda_{\rm c}^2c){A_{\rm an}\over\lambda_{\rm c}^2}{v_{\rm esc}\over c},
\qquad
n_{\rm e}\lambda_{\rm c}^2c=0.9\times10^{22}\,\rm s^{-1}.
\label{mult2}
\ee
With $A_{\rm an}/\lambda_{\rm c}^2=10^8$ and $v_{\rm esc}/c=0.1$, a total of $10^8$ such LTACRs would give $\,{\dot{\!N}}=10^{37}\,\rm s^{-1}$, consistent with what is required.

These estimates are subject to uncertainty and flexibility, associated with the location of the acceleration region relative to the chromosphere, and hence to the electron density $n_{\rm e}$, the skin depth ($\lambda_{\rm c}\propto1/n_{\rm e}^{1/2}$), and the length $\Delta R$ across the flux tube. Subject to these provisos, it appears that the model can account semi-quantitatively for the parameters $\,{\dot{\!N}}>10^{36}\,\rm s^{-1}$ and $\varepsilon\approx10^4\,$eV involved in bulk energization.

\section{Discussion}
\label{s:discussion}

We comment on four points related to out model: the acceleration mechanism, the return current, current concentration, and density depletions and enhancements.

The acceleration mechanism proposed here depends explicitly on the development of anomalous resistivity. We identify the accelerating electric field as $E_\parallel=J_\parallel/\sigma_{\rm an}$, with $J_\parallel$ set equal to the threshold current density for the Buneman instability, and with the anomalous conductivity identified as $\sigma_{\rm an}=\varepsilon_0\omega_{\rm p}^2/\nu_{\rm eff}$. The model requires that the effective collision frequency and the length of the anomalously conducting region ($\ell_\parallel=\Psi_{\rm max}/E_\parallel$) be related by $\nu_{\rm eff}\ell_\parallel=v_{\rm A}$ \cite{H12}. In other proposed mechanisms, acceleration is attributed to the electric field $E_\parallel$ associated directly with the Alfv\'en wave. The mechanism assumed by \inlinecite{MF09} depends explicitly on the IAWs being pulsed, with the parallel potential causing reflection of some electrons as the pulse passes. The mechanism assumed by \inlinecite{BKB10} involves $E_\parallel$ in KAWs rather than IAWs. Fermi acceleration, as suggested by \inlinecite{LMS94}, is related to Landau damping of magnetoacoustic waves \cite{A81}, so that this model requires that the turbulent cascade produce a magnetoacoustic component before it can accelerate electrons \cite{LM06}. The acceleration mechanism we propose is confined to LTACRs. Within a LTACR, the parallel potential is that implied by Ohm's law, and the energy inflow into a LTACR occurs both along and across field lines. The electric field $E_\parallel^+$ in the wave is essential for the cross-field inflow, and $\Psi^+$ in the wave determines the maximum parallel potential available along a LTACR.

The statistically large number ($10^8$) of LTACRs required leads to a simplification, due to the model depending more on the statistical distribution of LTACRs than on the (poorly-determined) detailed properties of individual LTACRs. We suggest a statistical model in which each LTACR is a current sheet of thickness (along the $x$-axis) of order a skin depth and extends across the flux tube (along the $y$-axis). In such a model, a weaker return current flows in the regions between the current sheets, and resupplies electrons to the acceleration region at the rate accelerated electrons escape from this region.

An essential requirement of our model is a mechanism that concentrates the current into regions of very high local current density. Possible mechanisms were discussed inconclusively by \inlinecite{H12}, and we do not repeat his arguments here. We note the confinement of Alfv\'en waves to small-scale current threads has been assumed in a possible explanation for coronal heating \cite{CVG10}. In our model \cite{MW13} there is an unrelated requirement for current concentration: the power transported by the Alfv\'en waves increases as the current associated with the wave increases. Assuming that the power release is driven towards a maximum, this necessarily involves the current becoming strongly concentrated, and favors any mechanism that tends to increase the current density in the resulting current sheets.

We note a general feature of the model whose implications require further investigation: the current flow necessarily results in density depletions and enhancements. This results from the properties of the current loops: electrons carry the current along field lines, and ions carry the (polarization) current across field lines. In any region where an upgoing current connects to a cross-field current there is a net outflow of particles, and in any region where a downgoing current connects to a cross-field current there is a net inflow of particles. Regions of charge depletion, referred to as cavities, are a notable feature of the auroral (upward-current) acceleration region (\opencite{Cetal03}, \citeyear{Cetal07}). Such current-driven depletions and enhancements are likely to play a significant role in a more detailed model, but we do not discuss these effects further here.

\section{Conclusions}
\label{s:conclusions}
The ideas discussed in this article indicate how the long-standing number problem \cite{MB89} might be solved. A new idea that we introduce is that the microphysics that determines the dissipation of the Alfv\'en waves by acceleration of electrons also constrains the global electrodynamics. Specifically, we show that there is a maximum parallel potential $\Psi_{\rm max}$, determined by the properties of IAWs together with the maximum current density that can flow without exciting anomalous resistivity. We argue that the existence of $\Psi_{\rm max}$ plays two important roles in the model. First, it determines the maximum energy ($\varepsilon_{\rm max}=e\Psi_{\rm max}$) to which electrons are accelerated. Second, it restricts the perpendicular potential $\Phi^+$ in the Alfv\'en wave to less than about $\Psi_{\rm max}$. We argue that this forces the system to drive the cross-field current $I$ in the generator region to flow up and down along field lines multiple ($M$) times in order to achieve the required dissipation. Each Alfv\'en wave transports a potential $\Phi/M$ and involves a current $I$ confined to a thin current sheet, with the return current flowing in the regions between the sheets.

The development of anomalous resistivity plays an essential role. We assume that bulk energization of electrons occurs in local, transient, anomalously-conducting regions (LTACRs). The strong concentration of current into narrow channels, required to account for the energy transport by Alfv\'en waves, is assumed to exceed the threshold for the Buneman instability, leading to anomalous resistivity. The energy transported downward by the Alfv\'en wave flows into LTACRs in the acceleration region both along and across field lines, with the latter allowed due to $E_\parallel\ne0$ in IAWs. We propose an idealized model in which the direct current is confined to sheets several skin depths in thickness, separated by regions of return current. The model can account semi-quantitatively for the inferred bulk energization.

Our results are sufficiently encouraging to warrant a more detailed model.

\begin{acks}
We thank Neil Cramer for helpful advice on dispersive Alfv\'en waves.
\end{acks}

\appendix
\section{Wave Equation for KAWs}
\label{A-appendix}
In kinetic theory, the response of the electrons is described in terms of the plasma dispersion function with an argument $y=(\omega^2/2k_z^2V_{\rm e}^2)^{1/2}$. The cold-electron limit corresponds to $y\gg1$. The parallel component of the dielectric tensor then reduces to the cold-plasma form $K_{zz}=1-\omega_{\rm p}^2/\omega^2$, and this cold-plasma response leads to Equation (\ref{IAW4}). In the opposite limit ($y\ll1$) one has
\be
K_{zz}=1-{\omega_{\rm pi}^2\over\omega^2}+{1\over k_z^2\lambda_{\rm De}^2},
\label{KAW1}
\ee
where the ions are assumed cold, $\omega_{\rm pi}$ is the ion plasma frequency, and $\lambda_{\rm De}=V_{\rm e}/\omega_{\rm p}$ is the electron Debye length. The final term in Equation (\ref{KAW1}) corresponds to a quasi-static, one-dimensional screening, satisfying $(\partial^2/\partial z^2+1/\lambda_{\rm De}^2)\Psi=-\rho_{\rm ext}/\varepsilon_0$ for any extraneous charge density $\rho_{\rm ext}$. Making the replacements $\omega^2\to-\partial^2/\partial t^2$ and $k_z^2\to-\partial^2/\partial z^2$, in place of Equation (\ref{IAW3}) one finds
\be
{\partial^2\over\partial z^2}\left[
{1\over c^2}{\partial^2\over\partial t^2}
+{\omega_{\rm pi}^2\over c^2}\right]\Psi-{1\over\lambda_{\rm De}^2}
{1\over c^2}{\partial^2\Psi\over\partial t^2}
+{1\over v_0^2}{\partial^2\over\partial t^2}
\nabla_\perp^2\Phi=0.
\label{KAW2}
\ee
Neglecting the first term in Equation (\ref{KAW2}) gives
\be
\Psi=R_{\rm g}^2\nabla_\perp^2\Phi,
\qquad
R_{\rm g}^2=\lambda_{\rm De}^2{c^2\over v_0^2}=\lambda_{\rm c}^2{V_{\rm e}^2\over v_0^2}.
\label{KAW3}
\ee
Combining Equations (\ref{we1}) and (\ref{KAW3}) one obtains the wave equation for KAWs in a form analogous to Equation (\ref{IAW5}):
\be
\left[{1\over v_0^2}{\partial^2\over\partial t^2}-(1-R_{\rm g}^2\nabla_\perp^2){\partial^2\over\partial z^2}\right]
\left(
\begin{array}{c}\Phi\\
\Psi
\end{array}
\right)
=0.
\label{KAW4}
\ee
After Fourier transforming, Equations (\ref{2pot}), (\ref{KAW3}), and (\ref{KAW4}) imply the wave properties in Equation (\ref{KAW}). The requirement $v_{\rm A}^2\ll V_{\rm e}^2$ renders KAWs of little interest in the present context.

More detailed treatments of KAWs are available \cite{H76,LL96,LS03,DWM09}. In particular, the value of the parameter $R_{\rm g}$ in Equation (\ref{KAW}) derived by \inlinecite{H76} is \cite{LL96}
\be
R_{\rm g}^2={V_{\rm i}^2\over\Omega_{\rm i}^2}\left({3\over4}+{T_{\rm e}\over T_{\rm i}}\right),
\label{Rg}
\ee
where $V_{\rm i}^2=T_{\rm i}/m_{\rm i}$ and $\Omega_{\rm i}=q_{\rm i}B/m_{\rm i}$ are the thermal speed and gyrofrequency, respectively, of ions of mass $m_{\rm i}$ and charge $q_{\rm i}$.

\section{Scalar and Vector Potentials}
\label{B-appendix}

\inlinecite{S00} derived the properties of IAWs using the conventional description of the electromagnetic field in terms of the scalar potential, $\phi$, and the vector potential, ${\bi A}$:
\be
{\bi E}=-{\bf\nabla}\phi-{\partial{\bi A}\over\partial t},
\qquad
\delta{\bi B}={\bf\nabla}\times{\bi A},
\label{B1}
\ee
where $\delta{\bi B}$ is the magnetic field in the wave. These authors gave an argument for assuming ${\bi A}_\perp=0$. (We interpret ${\bi A}_\perp=0$ as the gauge condition in the case where ${\bf\nabla}\times{\bi E}_\perp$ is set to zero, because it describes magnetoacoustic rather than Alfv\'en waves.) With ${\bi A}=A_z{\hat{\bi z}}$, Equation (\ref{B1}) becomes
\be
{\bi E}_\perp=-{\bf\nabla}_\perp\phi,
\quad
E_z=-{\partial\phi\over\partial z}-{\partial A_z\over\partial t},
\qquad
\delta{\bi B}={\bf\nabla}A_z\times{\hat{\bi z}}.
\label{B2}
\ee

An arbitrary gauge transformation involves an arbitrary function, $\psi$ say, with the new potentials given by
\be
\phi'=\phi+{\partial\psi\over\partial t},
\qquad
{\bi A}'={\bi A}-{\bf\nabla}\psi.
\label{B3}
\ee
The two-potential model follows by assuming $A'_z=0$, implying that $\psi$ is determined by $\partial\psi/\partial z=A_z$. One then finds
\be
\Phi=\phi,
\qquad
\Psi=\phi+\int dz\,{\partial A_z\over\partial t},
\qquad
{\bi A}'_\perp=-{\bf\nabla}_\perp\int dz\,A_z.
\label{B4}
\ee
We conclude that the two descriptions are equivalent.

\end{article}
\end{document}